\documentclass{article}
\usepackage{graphicx} 
\usepackage{color}
\newcommand{\blue}{\textcolor{black}}

\usepackage{tikz,pgfplots}
\usepackage{float}
\usepackage{amsmath,amsthm,amsfonts}
\usepackage[normalem]{ulem} 
\usepackage{xcolor}
\usepackage{cancel}

\usepackage{svg}

\newtheorem{lemma}{Lemma}
\theoremstyle{definition}

\theoremstyle{plain} 
\newtheorem{proposition}{Proposition}
\newcommand{\RR}{\mathbb{R}}
\newcommand{\EE}{\mathbb{E}}
\newcommand{\PP}{\mathbb{P}}
\title{Counting rankings of tree-child networks}
\author{Qiang Zhang and Mike Steel}

\begin{document}
\maketitle
\begin{center}
{\em Biomathematics Research Center, \\ School of Mathematics and Statistics, \\University of Canterbury, Christchurch, New Zealand}
\end{center}

\begin{abstract}
\noindent Rooted phylogenetic networks allow biologists to represent  evolutionary relationships between present-day species by revealing ancestral speciation and hybridization events. A convenient and well-studied class of such networks are `tree-child networks' and a `ranking' of such a network is a temporal ordering of  the ancestral speciation and hybridization events. In this short note, we investigate the question of counting such rankings on any given binary (or semi-binary) tree-child network. We also investigate the relationship between rankable tree-child networks and the class of `normal' networks.  Finally, we provide an explicit asymptotic expression for the expected number of rankings of a tree-child network chosen uniformly at random. 
\end{abstract}


\section{Introduction}

Rooted phylogenetic networks provide an effective model for biologists to represent the relationship between present-day species and their common ancestor through speciation and hybridization events \cite{hus10, mar14}.  Tree-child networks are a class of phylogenetic networks where each ancestral species has at least one path to the present via speciation events \cite{car08}. For some tree-child networks, it is possible to impose a time-stamp on each species in such a way that (i)  earlier species are assigned earlier time-stamps than their non-hybrid descendants, and (ii) hybrid species are assigned the same time stamp as their parents.  This assignment of time-stamps gives rise to a discrete temporal `ranking' of the vertices of the network.  This leads to some natural questions, such as:  `Does a given tree-child network $N$ have a temporal ranking?', `If so, how many different temporal rankings does $N$ have?', and `What is the average number of temporal rankings of a tree-child network chosen uniformly at random?'.

The answer to the first question can be no for certain tree-child networks \cite{bar06}, and in this paper, we further investigate the relationship between the existence of a ranking and the class of `normal' networks. We then introduce a new method to address the second question (i.e., to count the number of temporal rankings of any given separated tree-child network). Finally, we address the third question by deriving an asymptotic expression for the expected number of rankings of a tree-child network chosen uniformly at random.

\subsection{Definitions and notation}
\label{defsec}

A \textit{simple directed acyclic graph} is a directed graph with no directed cycles, no self-loops, and no multiple arcs between the same ordered pair of vertices. Let $D=(V,A)$ be a simple directed acyclic graph. $D$ is connected if there is a path (ignoring arc directions) between any two vertices. A \textit{descendant} of a vertex $v$ is any vertex $u$ that can be  reached  by  following  a  directed  path (possibly reduced to a single vertex)  from  $v$, denoted as $v \preceq u$.  We write $v \prec u$ if $v \preceq u$ and $v \neq u$.
A \textit{topological ordering} of $D$ is an ordering of all vertices $\tau=(v_1,v_2,\cdots, v_n)$ with the property that if there is a directed path from $v_i$ to $v_j$, then $i<j$. Define the \textit{index} induced by $\tau$ of a vertex $v$ by $\text{ind}_{\tau}(v_i)=i$. Note that a topological ordering is a total ordering of all vertices (i.e., any two vertices are comparable). Let $\delta(D)$ denote the number of topological orderings of $D$, and since $D$ is acyclic, we have  $\delta(D)\geq 1$ (e.g., by Proposition 1.4.3 of \cite{ban01}).

A \textit{rooted network} is a connected simple directed acyclic graph $N=(V,A)$ such that each vertex is either
\begin{itemize}
    \item a \textit{root vertex} of in-degree 0 and out-degree at least 2;
    \item  a vertex of in-degree 1 and out-degree at least 2;
    \item  a \textit{reticulate vertex} of in-degree $> 1$;
    \item  a \textit{leaf} of in-degree 1 and out-degree 0.
\end{itemize}

A \textit{tree vertex} is a vertex that is not a reticulate vertex. A \textcolor{black}{\textit{tree-to-tree vertex}} is a tree vertex that has tree vertices as children. A leaf is not a tree-to-tree vertex because it does not have any children. An \textit{internal vertex} is any vertex of out-degree $> 0$. An arc is a \textit{tree arc} if it ends at a tree vertex; otherwise, the arc is a \textit{reticulation arc}.  The \textit{out-degree} and \textit{in-degree} of any vertex $v$ are denoted $d^{+}(v)$ and $d^{-}(v)$, respectively.

A network is  \textit{separated} if all its reticulate vertices have out-degree 1. 

 \subsection{Phylogenetic networks}
A \textit{phylogenetic network} on a set of $X$ of distinctly labeled species is a rooted network $N=(V,A)$ such that $X=\{v \in V: d^{+}(v)=0, d^{-}(v)=1\}$ is a set of leaves.
A \textit{phylogenetic tree} is a phylogenetic network that has no reticulate vertices.

A {\em semi-binary} phylogenetic network is a separated network which has the properties that (i) each non-leaf tree vertex has out-degree $\geq 2$ and (ii) each reticulate vertex has \textcolor{black}{in-degree 2}.

A {\em binary} phylogenetic network is a separated phylogenetic network that has the property that each non-leaf tree vertex has out-degree 2 and each reticulate vertex has in-degree 2.

A phylogenetic network is \textit{non-binary} if it is a non-separated network (i.e., there is at least one reticulate vertex with out-degree $> 1$).

A \textit{tree-child network} is a phylogenetic network that has the property that each non-leaf vertex has a child that is a tree vertex. 

A \textit{normal} network is a tree-child network $N$ with the additional property that if $v_1,\cdots, v_k$ is a directed path in $N$ from $v_1$ to $v_k$ and $ k  >  2$, then $ (v_1   , v_k)$ is not an arc in $N$ (i.e., there are no `short-cut' arcs).

Let  $N=(V,A)$ be a \textcolor{black}{phylogenetic} network. Define a relation $\mathrel{R}$ on the set $\overset{\circ}{V}$ of internal vertices of
$N$ by:
$u\mathrel{R}v \Leftrightarrow u = v$, or $u$ and $v$ are linked \textcolor{black}{only by reticulation arc(s) if we ignore the direction of arc(s)}.

The proof of the following result is straightforward and provided in the Appendix.

\begin{lemma}\label{proseq}
    If $N=(V,A)$ is a \textcolor{black}{phylogenetic} network, $\mathrel{R}$ is an equivalence relation on $\overset{\circ}{V}$.
\end{lemma}

We call the equivalence classes of $\mathrel{R}$ the \textit{events} of $N$ and write $\bar{u}$ for the equivalence class of a vertex $u$.
\begin{itemize}
    \item Either $\bar{u}  =  \{u\}$, in which case $\bar{u}$  is  called  a {\em branching event}; or
\item $\bar{u}$  has at least three elements, and $\bar{u}$  is called  a  {\em reticulation  event}.
\end{itemize}

Given a \textcolor{black}{phylogenetic network} $N=(V,A)$, let 
\begin{equation}
\label{eqdN}
 dN=\{\bar{v} :v\in \overset{\circ}{V}, v \ \text{is a reticulate vertex or a \textcolor{black}{tree-to-tree vertex}}\}  
\end{equation}
be the set of equivalence classes of $\mathrel{R}$.

For two events $\bar{u}, \bar{v}$ of $N$, we say that $\bar{u}$ is descendant of $\bar{v}$ or $\bar{v}$ is ancestor of $\bar{u}$, denoted as $\bar{v}\preceq \bar{u}$, if there exist $v\in\bar{v}$ and $u\in\bar{u}$ such that $v\preceq u$ holds. We say that two events $\bar{v}$, $\bar{u}$ are $\preceq$-\textit{comparable} if $\bar{v}\preceq \bar{u}$ or $\bar{u}\preceq \bar{v}$. Note that $\bar{u}\preceq\bar{u}$ trivially always holds for any event $\bar{u}$.

A phylogenetic network \textcolor{black}{$N=(V,A)$} is said to have a {\em temporal labelling} if there is a function $t  :  V    \to  \RR^{\geq0}$ for which the following two properties hold:
\begin{itemize}
    \item T1: if  $(u, v )$  is  a  reticulation  arc,  then  $t(u)  =  t(v )$;
    \item T2: if  $(u, v )$  is  a  tree  arc,  then  $t(u)  <  t(v )$.
\end{itemize}

If a phylogenetic network \textcolor{black}{$N=(V,A)$} has a temporal labelling, then there is a function $r$ (called a {\em ranking}) taking values from the set $\{0,1,2,\cdots, \textcolor{black}{e_{N}-1}\}$ \textcolor{black}{where $e_N$ is the number 
of events of $N$} with the root vertex assigned rank 0  (i.e., $r(\rho) = 0$) and satisfying the following property: for each internal vertex $v$, if $(u, v )$ is a tree arc, then $r(u) < r(v)$, and if $v$ is a reticulate vertex, then $v$ has the same rank as all \textcolor{black}{internal vertices that are linked by reticulation arcs}. \textcolor{black}{This is the only case where equal ranking occurs.} Since all vertices \(u \in \bar{u}\) have the same rank, we define
\[
r(\bar{u}) := r(u) \quad \text{for any } u \in \bar{u}.
\] In the context of biology, rankings indicate a possible historical interpretation of evolution with the conditions that all ancestral species (along with the hybrid they formed) involved in a hybridization event must have been extant at the same time (T1) and ancestral speciation implies a passage of time (T2) \cite{kon22}. In phylogenetics, leaves are generally regarded as present taxa/species, while internal vertices represent ancestral species. Therefore, rankings  generally refer to internal vertices.

A network $N$ is \textit{rankable} if it has at least one ranking. Two rankings $r_1, r_2$ of $N$ are \textit{distinct} if there exists at least one vertex $v$ such that $r_1(v)\neq r_2(v)$. Note that a rankable network may have multiple distinct rankings. Given a network $N$, let $\psi(N)$ denote the number of rankings of $N$.

\begin{figure}[H]
\centering
\includegraphics[scale=0.75]{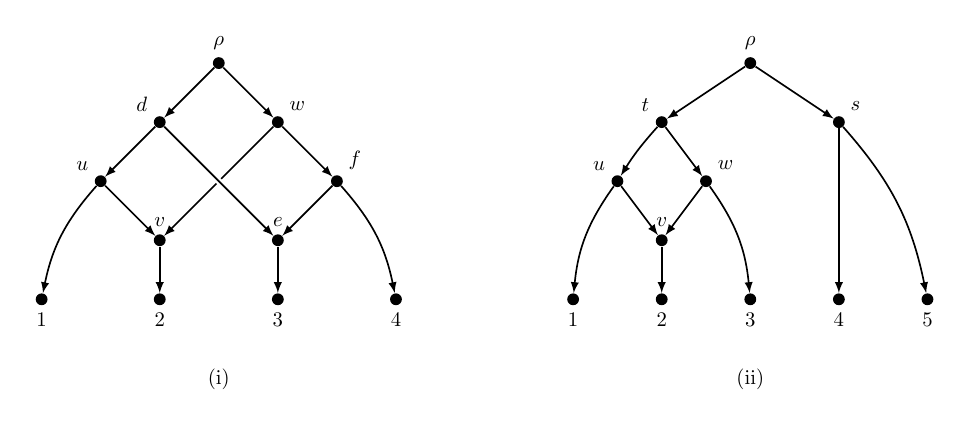}
\caption{(i) A binary normal network which is not rankable. (ii) A binary rankable normal network with three distinct rankings.}
\label{fig1}
\end{figure}

Fig.~\ref{fig1}(i) is an example of a binary normal network that has no ranking.  To see why, suppose that a ranking function $r$ exists. We would then have $r(u)=r(v)=r(w)=t_1$, and $r(d)=r(e)=r(f)=t_2$; in addition, we have $r(d)=t_2<r(u)=t_1$ (because $(d,u)$ is a tree arc) and $r(w)=t_1<r(f)=t_2$ (because $(w,f)$ is a tree arc). These last two inequalities $t_2<t_1<t_2$ provide a contradiction. On the other hand, Fig. \ref{fig1}(ii) is a binary rankable normal network and has three distinct rankings (i.e., we can let $r(s)<r(t)$ or $r(t)<r(s)<r(u)=r(v)=r(w)$, or $r(s)>r(u)=r(v)=r(w)$).

\subsection{Outline of results}
 It is known that every rankable binary tree-child network is normal; however, this does not extend to non-binary networks (we give a counterexample). Nevertheless, we show that \textcolor{black}{separated} tree-child networks that are rankable are normal.

In Section~\ref{sec2}, we investigate a transformation to facilitate  the enumeration of the rankings of any (binary or semi-binary) tree-child network (Proposition~\ref{prorank}),  thereby addressing a question posed at the end of Section 1.2 of \cite{bie22}.

Finally, we consider the expected number of rankings of a randomly-sampled binary tree-child network with $n$ leaves and $k$ reticulate vertices. \blue{We show this number is asymptotically of the form $\frac{1}{4^k} \cdot f(n)$ as $n$ grows.}

\section{Rankable \textcolor{black}{separated} tree-child networks are normal}
\label{sec1}

Although every binary rankable tree-child network is normal (see e.g., \cite{ste16} Prop. 10.12),  non-binary \textcolor{black}{rankable} tree-child networks can fail to be normal, as we show shortly.  Nevertheless, we also establish that every rankable \textcolor{black}{separated} tree-child network is normal. 
\begin{figure}[htb]
\centering
\includegraphics[scale=0.8]{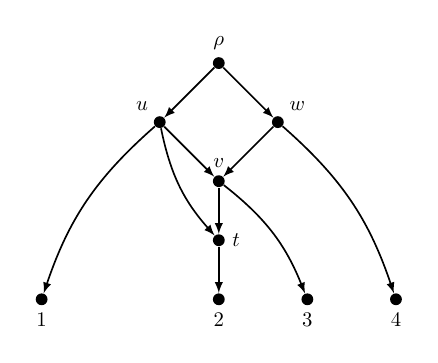}
\caption{A non-binary tree-child network which has a temporal ordering but is not a normal network.}
\label{fig2}
\end{figure}

The network in Fig.~\ref{fig2} is a non-binary tree-child network (i.e., $d^{+}(v)=d^{-}(v)=2$) which has a temporal ordering but is not normal (since $(u,t)$ is a short-cut arc). Note that $v$ and $t$ are both reticulate vertices. Although $t$ is a child of $v$, the network is still a tree-child network because $v$ has another leaf child (labelled as 3). We can temporally label all the vertices as follows: $r(\rho)=t_0=0, r(u)=r(w)=r(v)=r(t)=t_1$ and all the leaves have temporal label of $t_2$ such that $t_0 <t_1<t_2$.

To prove the first result of the paper, we begin with the following lemma.

\begin{lemma}\label{directed path}
    Given a rooted network $N$ and a directed path $v_1,v_2,\cdots,v_n$ of $N$ and $n \geq 3$, if  $N$ has a temporal labelling, then $r(v_1)\leq r(v_n)$ \textcolor{black}{for any ranking function $r$} and internal vertex $v_i$, $i\in\{1,2,\cdots,n\}$. 
\end{lemma} 

\begin{proof}
By definition of the ranking function, $r(v_i) \leq r(v_{i+1})$
for every arc $(v_i, v_{i+1})$ on the directed path $v_1,v_2,\cdots,v_n$.
Therefore, $r(v_1) \leq r(v_n)$.
\end{proof}

\begin{proposition}\label{separated}
If $N$ is a \textcolor{black}{separated} tree-child network that has a temporal ordering, then $N$ is normal.
\end{proposition}

\begin{proof}
    We provide a proof by contradiction. Suppose that $N$ is a \textcolor{black}{separated} tree-child network that has a temporal ordering and $N$ is not normal. $N$ is temporal, so it has at least one ranking $r$. Then $N$ has a directed path $v_1,v_2,\cdots,v_k$ such that $k \geq 3$ and $(v_1,v_k)$ is an arc. \textcolor{black}{Note that $v_k$ is a reticulate vertex such that $v_1$ and $v_{k-1}$ are parents of $v_k$; let $r(v_1)=r(v_{k-1})=r(v_k)=t_1$. Consider the vertex $v_{k-1}$. If $v_{k-1}$ is a reticulate vertex, then $v_{k-1}$ has exactly one child $v_k$, which is a reticulate vertex, and thus $N$ is not a tree-child network, which is a contradiction. Therefore, $v_{k-1}$ is a tree vertex and $r(v_{k-2})<r(v_{k-1})=t_1$. In addition, $v_1,\cdots,v_{k-1}$ is a directed path and, by Lemma \ref{directed path}, $r(v_1)=t_1\leq r(v_{k-2})$, which is a contradiction.}
    
\end{proof}

\section{Counting the rankings of (separated) tree-child networks}
\label{sec2}
Given a rooted tree $T=(V,A)$,  a standard result in enumerative combinatorics (e.g. \cite{knu97}) is the following:
\begin{equation}
    \label{numb}
    \delta(T)=\frac{\vert V\vert !}{\prod_{v \in V}\lambda(v)},
\end{equation} where $\lambda(v)=\vert\{u\in V: v \preceq u\} \vert$ and $\preceq$ is the partial order defined at the start of Section~\ref{defsec}.

Here we note that \textcolor{black}{the number of events (as defined in Section 1.2) of $T$ is just the number of internal vertices of $T$}, and counting rankings of $T$ is equivalent to counting topological orderings of the tree $T'=(V',A')$ obtained from $T$ by deleting leaves and their incident arcs. Therefore, $$\psi(T)=\delta(T')=\frac{\vert V'\vert !}{\prod_{v \in V'}\lambda(v)}.$$ However, for networks with reticulate vertices, the two concepts are different, and counting topological orderings is known to be \#P hard for general networks \cite{bri91}.

Consider a separated tree-child network $N=(V,A)$ with at least one reticulate vertex. We obtain an associated directed graph $\Psi(N)$ as follows. The vertex set of $\Psi(N)$ is $dN$, and for any two equivalence classes $\bar{u}$ and $\bar{v}$ of $dN$, $\bar{u}$ and $\bar{v}$ are joined by an arc $(\bar{u}, \bar{v})$ if there exist $t\in \bar{u}$ and $s\in \bar{v}$ such that $(t,s)$ is a tree arc of $N$ \cite{bar06}.

We denote the result of applying the above operations by writing 
$$\Psi(N)= \Tilde{N}.$$

\begin{figure}[htb]
\centering
\includegraphics[scale=1.0]{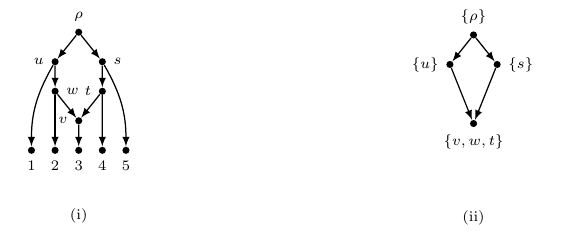}
\caption{(i) A separated tree-child network $N$ with five leaves and one reticulate vertex. (ii) The associated directed network $\Psi(N)$. }
\label{fig3}
\end{figure}

In the following, we will show that the number of rankings of $N$ is equal to the number of topological orderings of $\Psi(N)$, and we illustrate this on a few examples. As seen in Fig.~\ref{fig3}, (i) is a separated tree-child network $N$ with five leaves and one reticulate vertex. The associated directed graph $\Psi(N)$ is shown in (ii) with vertex set of $\{\{\rho\},\{u\},\{s\},\{v,w,t\}\}$.
 
Note that $N$ has two rankings (we can either let $r(u)<r(s)$ or $r(u)>r(s)$), and it is clear to see that $\Psi(N)$ has two topological orderings.

\begin{figure}[htb]
\centering
\includegraphics[scale=1.0]{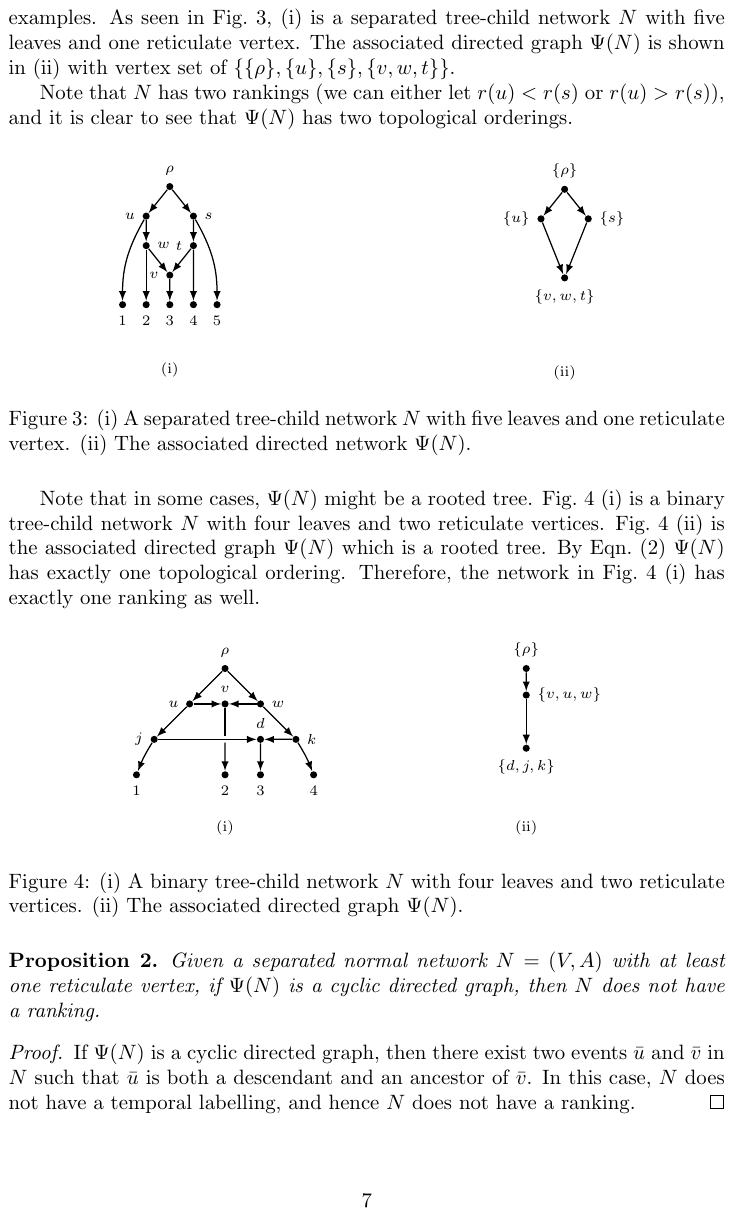}
\caption{ (i) A binary tree-child network $N$ with four leaves and two reticulate vertices. (ii) The associated directed graph $\Psi(N)$. }
\label{fig4}
\end{figure}

Note that in some cases, $\Psi(N)$ might be a rooted tree. Fig.~\ref{fig4}(i) is a binary tree-child network $N$ with four leaves and two reticulate vertices. Fig.~\ref{fig4}(ii) is the associated directed graph $\Psi(N)$ which is a rooted tree. By Eqn.~(\ref{numb}) $\Psi(N)$ has exactly one topological ordering. Therefore, the network in Fig.~\ref{fig4} (i) has exactly one ranking as well.

\begin{proposition}\label{normal not ranking}
Given a separated normal network $N=(V,A)$ with at least one reticulate vertex, if $\Psi(N)$ is a cyclic directed graph, then $N$ does not have a ranking.
\end{proposition}
\begin{proof}
    If $\Psi(N)$ is a cyclic directed graph, then there exist two events $\bar{u}$ and $\bar{v}$ in $N$ such that $\bar{u}$ is both a descendant and an ancestor of $\bar{v}$. In this case, $N$ does not have a temporal labelling, and hence $N$ does not have a ranking.
\end{proof}

\begin{figure}[htb]
\centering
\includegraphics[scale=1.0]{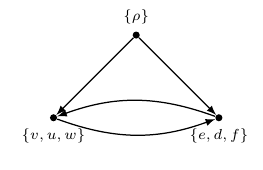}
\caption{The associated directed graph $\Psi(N)$ of the network $N$ (in Fig. \ref{fig1} (i)) has a cycle.}
\label{fig5}
\end{figure}

As an example of Proposition~\ref{normal not ranking}, the separated normal network $N$ in Fig. \ref{fig1} (i) does not have a ranking. The resulting directed graph in Fig. \ref{fig5} after Step (1) has been applied has a cycle which means that there are two distinct events $\bar{v}$ and $\bar{e}$ such that $\bar{v} \prec \bar{e}$ and $\bar{e} \prec \bar{v}$.

\begin{proposition}\label{prorank}
    Given a rankable separated tree-child network $N=(V,A)$, the number of possible rankings of $N$ equals the number of topological orderings of $\Psi(N)=(\Tilde{V}, \Tilde{A})$.  If $\Psi(N)$ is a rooted tree, we have: 
    
 $$\psi(N)=\delta(\Psi(N))=\frac{\vert \Tilde{V}\vert !}{\prod_{v \in \Tilde{V}}\lambda(v)},$$
    where $\lambda(v)$ is the number of vertices in $\Tilde{V}$ reachable from $v$ by a directed path (including $v$).
\end{proposition}

\begin{proof}
Let $\mathcal{K}(N)$ denote the set of rankings of $N$, and $\mathcal{T}(\Psi(N))$ the set of topological orderings of $\Psi(N)$.

Define a function
\[
f \colon \mathcal{K}(N) \to \mathcal{T}(\Psi(N)), \quad f(r) = \tau,
\]
where $\tau$ is the ordering of $\Psi(N)$ satisfying
\[
r(\bar{u}) = \operatorname{ind}_\tau(\bar{u}) - 1 \quad \text{for each event } \bar{u} \text{ of } N.
\]
Then for each $r \in \mathcal{K}(N)$, there is a unique $\tau \in \mathcal{T}(\Psi(N))$ sorting the vertices according to their ranks in $N$. Therefore, $f$ is well defined.

Next, we show that $f$ is a bijection. Suppose $r_1, r_2 \in \mathcal{K}(N)$ are distinct but $f(r_1) = f(r_2) = \tau$. Then there exists an event $\bar{u}$ of $N$ with $r_1(\bar{u}) \neq r_2(\bar{u})$. However, by definition of $f$, we have
\[
r_1(\bar{u}) = \operatorname{ind}_\tau(\bar{u}) - 1 = r_2(\bar{u}),
\]
a contradiction. Therefore, $f$ is injective.

For each $\tau \in \mathcal{T}(\Psi(N))$, define $r \in \mathcal{K}(N)$ by
\[
r(\bar{u}) = \operatorname{ind}_\tau(\bar{u}) - 1 \quad \text{for all events } \bar{u} \text{ of } N.
\]
Then $f(r) = \tau$, so $f$ is surjective. Together with injectivity, this shows that $f$ is a bijection.
Hence, the number of rankings of $N$ is the number of topological orderings of $\Psi(N)$. 
 If $\Psi(N)$ is a rooted tree, the number of topological orderings of $\Psi(N)$, $\delta(\Psi(N))$, is given by Eqn.~(\ref{numb}):
    $$\delta(\Psi(N))=\frac{\vert \Tilde{V}\vert !}{\prod_{v \in \Tilde{V}}\lambda(v)},$$
    which is the number of rankings of the separated tree-child network $N$.
\end{proof}

\noindent {\bf Remarks:}
A phylogenetic network without any reticulate vertex is just a rooted tree. The number of rankings of a rooted binary tree $T=(V,A)$ with $n$ leaves is $$\frac{(n-1)!}{\prod_{v \in \Tilde{V}}\lambda(v)},$$ where $\Tilde{T}=(\Tilde{V}, \Tilde{A})$ is the rooted binary tree obtained by deleting all leaves of $T$, and $\lambda(v)$ is the number of vertices of $\Tilde{T}$ descended from $v$ (including $v$). The number of rankings of a rooted binary tree $T$ is just the number of topological orderings of $\Tilde{T}$, and the number of vertices of $\Tilde{T}$ is $n-1$.
In particular, certain rooted binary trees have exactly one ranking. More precisely a rooted binary tree has exactly one ranking if and only if it is a {\em caterpillar tree}, in which all the interior vertices form a directed path.

\section{The number of rankings of a random binary tree-child network}
\label{sec3}

Let $X_{n,k}$ be the random variable that describes the number of rankings of a binary tree-child network on leaf set $[n]=\{1, \ldots, n\}$ with $k$ reticulate vertices, chosen uniformly at random. Here $k$ is fixed, and we let $n$ grow.

The following result reveals that asymptotically (as $n$ grows) the expected number of rankings asymptotically splits into a product of two functions; one involving just $k$, the other just $n$.
Moreover, for any fixed $k$, it becomes increasingly certain that a random binary tree-child network will have at least one ranking as $n$ grows.

\begin{proposition}
For each fixed $k \geq 0$, as $n \rightarrow \infty$, the following hold:
\begin{itemize}
    \item[(i)] $$ \EE[X_{n,k}] \sim \frac{1}{4^k} \cdot 
 \frac{n!}{ \binom{2n-2}{n-1}}.$$
In particular, $\lim_{n \rightarrow \infty} \frac{\EE[X_{n,{k+1}}]}{\EE[X_{n,k}]}= \frac{1}{4}$,
for each $k \geq 1$.
    \item[(ii)] $\lim_{n \rightarrow \infty} \PP(X_{n,k}\geq 1) =1.$
\end{itemize}

\end{proposition}

\begin{proof}

{\em Part (i)} If a tree-child network is chosen uniformly at random, then:
\begin{equation}
\label{eqe}
    \EE[X_{n,k}] = \frac{RTCN(n,k)}{TCN(n,k)}.
\end{equation}
where 
 $RTCN(n,k)$ denotes the number of ranked tree-child networks on the leaf set $[n]$ with $k$ reticulate vertices (i.e.,  the number of ordered pairs $(T, r)$, where $T$ is a tree-child network, and $r$ is a ranking of the vertices of $T$) and 
$TCN(n,k)$ denotes the number of tree-child networks on the leaf set $[n]$ with $k$ reticulate vertices.

From  \cite{bie22} (Theorem 1), we have:
\begin{equation}
\label{exact_count}
RTCN(n,k)=
\begin{bmatrix}
n-1\\n-1-k
\end{bmatrix} \cdot \frac{n!(n-1)!}{2^{n-1}}\end{equation}
where 
$
\begin{bmatrix}
n-1\\n-1-k
\end{bmatrix}$
refers to the unsigned Stirling number of the first kind (i.e., the number of permutations on $n-1$ elements that have $n-1-k$ cycles). For $k$ fixed, and as $n \rightarrow \infty$, we have the following result from \cite{mos58} (Eqn. 1.6):
\begin{equation}
\label{aseq}
  \begin{bmatrix}
n-1\\n-1-k
\end{bmatrix} 
\sim \frac{(n-1-k)^{2k}}{2^k k!}  
\end{equation}
Note that the second term in Eqn. (\ref{exact_count}), namely, $\frac{n!(n-1)!}{2^{n-1}}$,  is the number of ranked rooted binary trees on leaf set $[n]$. 

Moreover, for any fixed values of $k$,  we have the following asymptotic equivalence as $n\rightarrow \infty$ from \cite{fuc24}:
\begin{equation}
TCN(n,k) \sim\frac{(2n^2)^k}{k!} r(n),
    \label{tcn}
\end{equation}
where 
\begin{equation}\label{tcn2}
r(n) =  \frac{(2n-2)!}{(n-1)!2^{n-1}}
\end{equation}
 is the  number of rooted binary phylogenetic trees on leaf set $[n]$.

Applying Eqns. (\ref{eqe}) -- (\ref{tcn2}), and noting that $(n-1-k)^{2k}/n^{2k} \rightarrow 1$ as $n\rightarrow \infty$ (for fixed $k$)  gives the claimed result.

{\em Part (ii)} This follows from results in \cite{fuc24}, which show that the proportion of binary tree-child networks with $k$ reticulate vertices and $n$ leaves that  have at least one ranking tends to 1 as $n \rightarrow \infty$.  
\end{proof}

\section{Concluding comments}

In this paper, we have shown that every rankable  separated tree-child networks is a normal network by establishing that for any rankable network $N$ with a directed path $v_1, v_2, \cdots, v_n$ ($n\geq 3$), the temporal labelling of $v_1$ is at most that of $v_n$. Furthermore, given a  separated tree-child network $N=(V,A)$, we can transform $N$ to a simpler network $\Psi(N)$. We have defined a function from the set of rankings of $N$ to the set of topological orderings of $\Psi(N)$ and shown that it is bijective.  Consequently, counting the number of rankings of $N$ is equivalent to counting the number of topological orderings of $\Psi(N)$, which is a standard problem in enumerative combinatorics \cite{knu97} and efficiently solvable when $\Psi(N)$ is a tree. A possible question for future work could be to determine whether the problem of counting the number of rankings of separated tree-child network is $\#P$-complete.

Finally, we have investigated the expected number of rankings of a tree-child network with $n$ leaves and $k$ reticulate vertices selected uniformly at random, revealing a curious and simple asymptotic factorization into the product of a term involving just $k$ and a term involving just $n$.  Describing the asymptotic distribution of  $X_{n,k}$ (suitably normalised) could be an interesting question for further work. 
\bibliographystyle{plain}
\bibliography{Zhang_Steel_26Jan2026_submission}

\section{Acknowledgements}  We thank the New Zealand Marsden Fund (23-UOC-003) for supporting this research. We also thank Michael Fuchs for some helpful comments, and the two anonymous reviewers for their valuable feedback.

\section{Appendix:  Proof of Lemma~\ref{proseq}}

    For any vertex $v \in \overset{\circ}{V}$, we have $v\mathrel{R}v$, so $\mathrel{R}$ is reflexive. Moreover, for any $v, u \in \overset{\circ}{V}$, if $v\mathrel{R}u$, then either $u = v$, or $u$ and $v$ are linked only by reticulation arc(s). Therefore, we also have $u\mathrel{R}v$, and thus $\mathrel{R}$ is symmetric. Thus, it remains to establish transitivity of $\mathrel{R}$. 
    
    Suppose that for $v, u, t \in \overset{\circ}{V}$, we have $v\mathrel{R}u$ and $u\mathrel{R}t$. We claim that $v \mathrel{R}t$.  This holds trivially if $u=v$,  while if $u \neq v$, $u$ and $v$ are linked by reticulation arcs. Given that  $u\mathrel{R}t$, if $u=t$, then $v \mathrel{R}t$; if $t$ and $u$ are linked by  reticulation arcs, then $u$ and $t$ are linked by reticulation arcs too. Thus, we have $v\mathrel{R}t$.

In summary, for any $v, u, t \in \overset{\circ}{V}$, if $v\mathrel{R}u$ and $u\mathrel{R}t$, then $v \mathrel{R}t$, thus $\mathrel{R}$ is transitive, and so $\mathrel{R}$  is an equivalence relation on $\overset{\circ}{V}$.

\end{document}